\documentclass[journal]{IEEEtran}
\usepackage{amsmath,amsfonts}
\usepackage{algorithmic}
\usepackage{algorithm}
\usepackage{array}
\usepackage[caption=false,font=normalsize,labelfont=sf,textfont=sf]{subfig}
\usepackage{textcomp}
\usepackage{stfloats}
\usepackage{url}
\usepackage{verbatim}
\usepackage{graphicx}
\usepackage{cite}
\usepackage{booktabs}
\usepackage{subfig}
\usepackage{float}
\usepackage{nomencl}
\usepackage[table]{xcolor}
\usepackage{multirow}
\usepackage{makecell}
\usepackage{diagbox}
\usepackage{bm}
\usepackage{balance}
\usepackage[colorlinks,
            linkcolor=black,
            anchorcolor=black,
            citecolor=black]{hyperref}
\usepackage[bottom]{footmisc}
\definecolor{mycolor_m}{RGB}{255,245,239} % 目录标题颜色
\definecolor{mycolor_c2}{RGB}{241,247,255} % 内容颜色

\definecolor{mycolor_r1}{RGB}{255,197,126} %综述列表题目颜色
\definecolor{mycolor_r2}{RGB}{255,173,71} %综述列表内容颜色
\definecolor{mycolor_r3}{RGB}{175,209,255} %综述列表内容颜色%导入自定义颜色包
\begin{document}

\title{A Simple and Extremely Efficient Predictive Control for Power Converters}
\author{
\vspace{-0mm}
Guangze Chen, Zhenbin Zhang,~\IEEEmembership{Senior Member,~IEEE}
\vspace{-8.5mm} 
}

% The paper headers
\markboth{Xxx}%
{Shell \MakeLowercase{\textit{et al.}}: A Sample Article Using IEEEtran.cls for IEEE Journals}

\maketitle

\begin{abstract}
Classical finite control set based model predictive control (FCS-MPC) reduces the optimal problems to an enumerated searching algorithm, which is very simple and effective to control power converters. However, it requires a large amount of enumeration operations, increasing its computational load and hardware costs. In this work, we propose a new and simple predictive control technique with extreme efficiency. The proposal directly selects the optimal vector via determined visual maps, abstained solely requiring a rearrangement of the cost function and a simple fitting law, without any enumeration. It has been validated under a lab-constructed power converter and a set of commercialized low-cost digital controllers. Experimental data confirm that the proposal achieves the same control performance as classical MPC, with significant computational burden reduction (up to \textit{88\% for one-step prediction of two-level converters}).
\end{abstract}


\section{Introduction}
\IEEEPARstart{P}{ower} converters, as fundamentals in electronic-based energy conversion systems, have been in continuous evolution since the second half of the 20th century, and significantly facilitate the techniques in almost every area~\cite{4663816}. 

%A whole power conversion The controller is a significant part of a power electronics system. It contributes dominantly, e.g., in terms of control-target tracking accuracy and disturbance-resisting capability. 

%一个完整的基于变流器的功率系统可以被抽象成三部分：1）器件、2）拓扑、3）控制。对于前两项而言，它们是变流系统的基础，其性能将直接影响变流系统的上限。更加先进的功率器件（更高功率密度以及开关频率），将为系统带来更加优异的性能（如体积，效率，电能质量等等）。控制是系统运转的大脑，对能否充分释放既有变流系统的性能具有重要意义，自上世纪以来受到了众多专家学者的广泛关注。

\begin{figure}[tbp]
	\centering
	\includegraphics[width=3.5in]{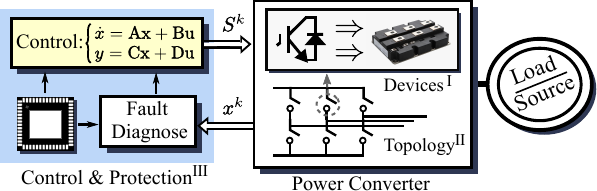}
	\caption{A typical converter-based energy conversion structure.}
	\label{fig_1}
\end{figure}

A typical converter-based energy conversion, well-known, is composed of three primary parts: 1) devices, 2) topologies, and 3) control \& protection (see, Fig.~\ref{fig_1}). On the one hand, advanced power devices can enhance system efficiency, power density, and upper performance limits, while innovative topologies optimize conversion processes, enabling greater flexibility and scalability. Protection ensures the safety and reliability of the system, while control is the key to unlocking the converter’s maximum potential within its physical limits.

History of controls in power electronics can be traced back to the middle of the 1950s, initially relying on manual or simple operations. With the rapid development of integrated circuits and digital controllers, more powerful control strategies have emerged, including sliding mode control~\cite{10488725}, adaptive control~\cite{9970754}, and model predictive control~\cite{4682711}, etc.

Model predictive control (MPC), standing out for its capability to manage multiple objectives with fast control dynamics, has been a very active research topic in the field of power electronics, covering countless applications. According to the continuity of the control signal, MPC can be mainly categorized into two types~\cite{6317184}: 1) continuous control set (CCS) MPC, and 2) finite control set (FCS) MPC.

CCS-MPC, which aims to globally optimize the desired targets over the prediction horizon, generates an optimal continuous control signal for the power converter. This signal typically requires a modulator to convert it into a switching sequence, which leads to better steady-state performance in the medium/high switching frequency. However, the derivation of the control signal is often complex, which typically involves the solution of a quadratic programming (QP) problem~\cite{10547342}. The complexity increases considerably as more optimized targets are integrated into the system.

FCS-MPC simplifies the optimization problem by exploiting the finite switching states of a specific power converter, reducing the problem to an enumerated search algorithm~\cite{4084698}. Due to its straightforward concept and robust capability to address multiple control objectives, FCS-MPC has garnered significant attention in a wide area~\cite{10366843, 10561621}. With the capability to directly identify the switching signals, FCS-MPC has a more powerful capability to handle the nonlinear switching characteristic than traditional modulator-based methods, which is obvious with low switching frequency or low carrier ratios; thereby, it is favorable for the high power applications with limited switching action in a fundamental~\cite{10665278,6744601}. However, the realization of such techniques usually enumerates all possible switching positions of the underlying power converter, resulting in a substantial computational burden and a considerable increase in the hardware cost~\cite{10572487}.

Reducing the computational load of FCS-MPC is a desirable topic since the method has been proposed. For the past works, the effort can be generally divided into two parts: 1) deadbeat type, 2) candidates reduction type. For the former, such techniques assume the system can reach the ideal position at the next sampling time, solving the optimal voltage vector, and directly outputting the switching pulse sequence through rounding. This approach is effective and widely used, but in multi-objective control scenarios, deriving the optimal voltage vector is challenging, even when only addressing switching frequency suppression. Although deadbeat predictive control based on modulation can manage the switching frequency, integrating multiple objectives remains difficult, and the method is limited by assumption (reach the ideal position at next sampling) and modulation delay effects, making it unsuitable for low switching frequency applications (\text{$<$}5kHz). For finding a more general way, the type of candidates reduction methods has arisen, which adopts several processes to reduce the enumerated candidates, thereby decreasing the computation load. E.g., in~\cite{7776884} the authors established a computationally efficient method based on an area classification concept. Through utilizing the ideal voltage vector solved by a deadbeat-type solver, the candidate states can be reduced into a pre-stored area (reduce the candidates from 27 to 12). The authors of~\cite{10432783} introduced a fast-optimization predictive control, which limits the candidate region to the near area of the last optimal vector, thereby reducing the numerous candidates of the modular multilevel converter to a much lower level. Similar concepts have been extended in several different topologies, e.g., five-phase three-level drives~\cite{10380325}, T-type three-level converters~\cite{10319837}, etc., and perform powerful capability in the computational reduction.
However, the improvement is not none sacrifices. The method in~\cite{7776884} will lead to a non-optimal switching penalty, while the method in~\cite{10432783} has a sacrifice on the dynamics. Apart from the candidate reduction, in~\cite{8974618} an intelligent-based technique was utilized to reduce the load. The method constructs an artificial neural network to imitate the behavior of FCS-MPC. Due to the fixed layers of such networks, the computational load is decoupled from the prediction horizon, thereby, achieving a significant reduction in multiple step prediction. The concept is interesting and soon extended to several applications~\cite{9424457}. However, this concept is limited to deal with one step prediction (the most common scenario of FCS-MPC), for which the load will oppositely increase.

In this work, a simple and efficient model predictive control technique is presented, referred to as direct mapping MPC (DM-MPC). Different from classical FCS-MPC, the proposal integrates a set of mapping laws to directly orient the optimal vector without any enumeration. Therefore, a significant reduction of the computational burden is achieved. Since the mapping laws are derived from hypothetical virtual errors, these laws are independent from the system, resulting in a potentially universalized solution for different control objectives, e.g., current, power (power grid), torque/flux (machines), etc. Furthermore, existing techniques for enhancing robustness or steady performance, e.g.,~\cite{7878550,7865950}, can be seamlessly integrated within this framework. The classical two-level power converter has been selected as the main topology for discussion. The reasons are its more generic structure and lower entry threshold. Major contributions include:
\begin{enumerate}
    \item A direct optimal oriented predictive control scheme is proposed that without any enumeration process, facilitating its implementation on low-costs microprocessors.
    \item The proposal substantially reduces the computational load around 88\% at no control performance sacrifice.
     \item The system information are integrated in the designed inputs, making the mapping law an universality in different objectives, e.g., current, power, torque/flux, etc.
\end{enumerate}

The rest is structured as follows. Sec. \uppercase\expandafter{\romannumeral2} revisits a simple power converter system and the classical FCS-MPC technique. Sec. \uppercase\expandafter{\romannumeral3} introduces the proposed DM-MPC technique. Sec. \uppercase\expandafter{\romannumeral4} illustrates the validation. While Sec. \uppercase\expandafter{\romannumeral5} concludes this paper.

\begin{figure}[tbp]
	\centering
	\includegraphics[width=3in]{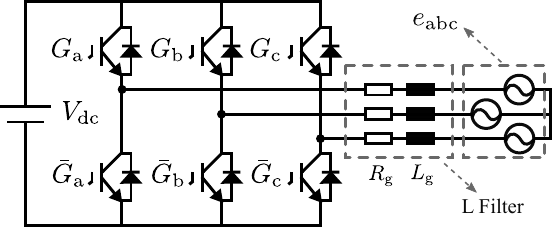}
	\caption{Topology of the two-level grid-tied power converter.}
	\label{fig_2}
     \vspace{-5mm}
\end{figure}

\section{System Modeling and Technique Review}

In this section, the discrete state-space model of the controlled two-level energy conversion system is firstly established, and the conventional FCS-MPC is reviewed\footnote{
Note that all quantities \text{$\bm{x}_{\alpha \beta}$} in the static coordinate and quantities \text{$\bm{x}_{\rm dq}$} are derived from \text{$\bm{x}_{abc}$} in the original coordinate through Clark and Park transformation, i.e.,
\begin{equation}
\begin{aligned}
& \text{Clarke:} 
& \mathbf{M}_{\mathrm{C}} = \frac{2}{3}
\begin{bmatrix}
   1 & -\frac{1}{2} & -\frac{1}{2} \\
   0 & \frac{\sqrt{3}}{2} & -\frac{\sqrt{3}}{2} \\
\end{bmatrix}
\Rightarrow
\bm{x}_{\alpha \beta}=\mathbf{M}_{\mathrm{C}}\bm{x}_{a b c}, \\
& \text{Park:} 
& \mathbf{M}_{\mathrm{P}} = 
\begin{bmatrix}
   \cos(\theta) & \sin(\theta)\\
   -\sin(\theta) & \cos(\theta)\\
\end{bmatrix}
\Rightarrow
\bm{x}_{\rm dq}=
\mathbf{M}_{\mathrm{P}}\bm{x}_{\alpha \beta}.
\end{aligned}
\nonumber
\end{equation}

In the following, all models are derived in a discrete format by applying the forward Euler method, i.e., \text{$\frac{d}{dt}x(t) \approx \frac{x_{[ k+1 ]}-x_{[ k ]}}{T_s}$}.
}.

\subsection{System Modeling}

Two-level power converter, as a basic topology in the modern power electronic based energy conversion systems, is taken as a modeling example. As depicted in Fig.~\ref{fig_2}, six switching tubes are integrated into the two-level converter, allowing for the generation of six active voltage vectors and two redundant zero states, i.e.,
\begin{equation}
\label{equ1}
\bm{u}_{abc}\in\mathcal{U}_{8}:=\{NNN,NNP,\ldots,PPN,PPP\},
\end{equation}
where P represents the positive state, indicating that the phase voltage of the converter is at a high level, i.e., \text{$G_{x}=1$, $\bar{G}_{x}=0$}. In contrast, N denotes the negative state, meaning the voltage is at a low level, i.e., \text{$G_{x}=0$, $\bar{G}_{x}=1$}. Assuming the voltage of the DC bus is \text{$V_{\rm dc}$}, the output voltage \text{$\bm{v}_{abc}$} can be described as

\begin{equation}
\label{equ2}
\bm{v}_{abc} = 
\begin{bmatrix}
v_{a}\\
v_{b}\\
v_{c}
\end{bmatrix} = \frac{V_{\rm dc}}{6}
\begin{bmatrix}
    2 & -1 & -1 \\
    -1 & 2 & -1 \\
    -1 & -1 & 2 \\
\end{bmatrix}\bm{u}_{abc}.
\end{equation}

The corresponding voltage \text{$\bm{v}_{\alpha \beta}$} in static coordinate can be directly obtained using the Clarke transformation. Based on Kirchhoff’s law, the discrete state-space model for the grid-tied conversion system can be established as 
\begin{equation}
\label{equ3}
    \bm{x}_{\varsigma}^{k+1}=\mathbf{A}_{\varsigma} \cdot \bm{x}_{\varsigma}^{k} + \mathbf{B}_{\varsigma}^{k} \cdot \bm{v}_{\varsigma}^{k} - \mathbf{B}_{\varsigma}^{k} \cdot \bm{e}_{\varsigma}^{k},
\end{equation}
where \text{$\varsigma \in \{\alpha \beta , \mathrm{abc} \}$}, representing the coordinate of the state-space model, and \text{$\bm{e}_{\varsigma}$} is the grid voltage. Take \text{$\alpha 
 \beta$} as an example, \text{$\mathbf{A}_{\varsigma}$} and \text{$\mathbf{B}_{\varsigma}$} are system matrices given as
\begin{equation}
\label{equ4}
   \mathbf{A}_{\varsigma}=
   \begin{bmatrix}
       1-\frac{T_\mathrm{s}R_\mathrm{g}}{L_\mathrm{g}} & 0 \\
       0 & 1-\frac{T_\mathrm{s}R_\mathrm{g}}{L_\mathrm{g}}
   \end{bmatrix},
   \mathbf{B}_{\varsigma}=
   \begin{bmatrix}
       \frac{T_\mathrm{s}}{L_\mathrm{g}} & 0 \\
       0 & \frac{T_\mathrm{s}}{L_\mathrm{g}}
   \end{bmatrix},
\end{equation}
where \text{$R_\mathrm{g}$} and \text{$L_\mathrm{g}$} are the parameters of the grid-tied filters.

\begin{figure}[tbp]
	\centering
	\includegraphics[width=3.5in]{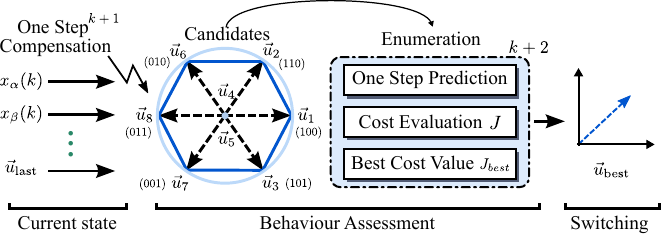}
	\caption{Control diagram of classical FCS-MPC.}
	\label{fig_3}
     \vspace{-5mm}
\end{figure}

\subsection{Classical FCS-MPC}

FCS-MPC, with simple implementation logic but powerful control capability, is widespread in the control of power converters~\cite{7733147}. Its deployment often contains three steps (see, Fig.~\ref{fig_3}): 1) state prediction, 2) cost evaluation, and 3) one-step compensation.

\textit{1) State Prediction}: Based on Eq.~\ref{equ3}, the current prediction model of \text{$\alpha\beta$-axes} can be expressed as follows:
\begin{equation}
\label{equ5}
\begin{cases}
i_{\alpha}^{k+1}=(1-\frac{R_{\rm g}T_{\rm s}}{L_{\rm g}})i_{\alpha}^{k}+\frac{T_{\rm s}}{L_{\rm g}}(v_{\alpha}^{k}-e_{\alpha}^{k}) \\
i_{\beta}^{k+1}=(1-\frac{R_{\rm g}T_{\rm s}}{L_{\rm g}})i_{\beta}^{k}+\frac{T_{\rm s}}{L_{\rm g}}(v_{\beta}^{k}-e_{\beta}^{k})
\end{cases}
\end{equation}
where \text{$\bm{u}_{abc}$} represents the switching combination, and \text{$i^{k+1}$} denotes the system currents in the future \text{$k+1$} instant.

\textit{2) Cost Evaluation}: To assess the effect of each candidate vector \text{$\bm{u}_{abc}$} on the desired control targets, a cost function needs to be established. The cost function can be designed flexibly to include multiple targets based on specific requirements, e.g., power~\cite{7055326}, current, torque/flux~\cite{5722040}. Take current tracking as an example, the cost function can be expressed as
\begin{equation}
\label{equ6}
\begin{aligned}
    J_\mathrm{g}= \| {i}_{\alpha}^{*} - {i}_{\alpha}^{k+1} \|^{2} + \| {i}_{\beta}^{*} - {i}_{\beta}^{k+1} \|^{2},
\end{aligned}
\end{equation}

\textit{3) One Step Compensation}:
Due to the computational constraints of the microprocessor, the selected voltage vector will not be output until the computation is complete, which causes considerable deterioration in control performance. Thereby, to compensate for the above delay, a one-step prediction is often utilized: the algorithm first outputs the optimal vector calculated in the previous period, and then uses this output vector to add one-step compensation before state prediction (see~\cite{5771558} for more details). This approach can mitigate the impact of the computational delay.

FCS-MPC is a competitive alternative for addressing complex control targets in power electronics and has been widely utilized. However, its evaluation process, typically requiring a great number of enumerations, imposes a considerable computational burden, which significantly increases the hardware costs. This is one of the significant drawbacks during the competition with other control methods in industrial applications, as production costs are crucial for commercial products.

\begin{figure*}[!tbp]
	\centering
	\includegraphics[width=6.3in]{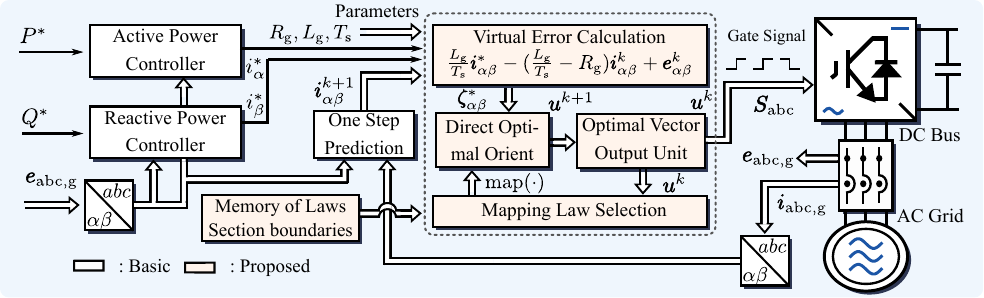}
	\caption{Overall diagram of the proposed direct mapping MPC.}
	\label{fig_4}
    \vspace{-5mm}
\end{figure*}

\begin{figure}[!hbp]
	\centering
	\includegraphics[width=3in]{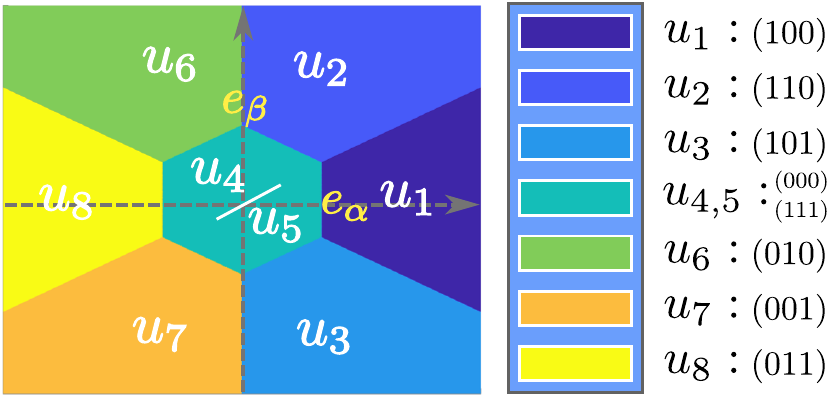}
	\caption{The visual mapping law under basic control targets.}
	\label{fig_5}
     \vspace{-5mm}
\end{figure}

\section{Proposed Direct-Mapping MPC Strategy}
The inspired concept of the proposed DM-MPC is that "no matter how FCS-MPC operates, if the state and cost function are identified, the best switching behavior is only/already determined". Thereby, the enumerated evaluation can be avoided. The only question is how to find such a mapping law, to map the measured data with the control demands to the optimal switching vector. In this section, we desire the mapping law with the following features:
\begin{enumerate}
    \item It has a great universality for different control targets, e.g., current/power tracking, torque/flux control.
    \item It has a capability to separated to system parameters, i.e., parameter related items are integrated in the input.
    \item Its derivation is simple and determined, without any uncertainty coefficients or hard training process.
\end{enumerate}

The overall diagram of the proposal is illustrated in Fig.~\ref{fig_4}. The detailed processes are introduced in the following.

\subsection{Virtual Error Establish}
To mitigate the influence of different systems on mapping laws, a set of virtual errors is first transferred, i.e., the original cost function is rebuilt as 
\begin{equation}
\label{equ7}
\begin{aligned}
    J_\mathrm{g}= \| {\zeta}_{\alpha}^{*} - {u}_{\alpha}^{k} \|^{2} + \| {\zeta}_{\beta}^{*} - {u}_{\beta}^{k} \|^{2},
\end{aligned}
\end{equation}
where \text{${\zeta}_{\alpha}$}, \text{${\zeta}_{\beta}$} are the transferred virtual errors and \text{${u}_{\alpha}$}, \text{${u}_{\beta}$} are the standard voltage vectors in the static coordinate, i.e., \text{$\bm{u}_{\alpha\beta} = \mathbf{M}_{\mathrm{C}}\bm{u}_{\rm abc}$}, directly relating to the switching vector.

Take the constructed two-level system in Sec.~\uppercase\expandafter{\romannumeral2} as an example, the first item in the cost function, i.e., \text{$({i}_{\alpha}^{*} - {i}_{\alpha}^{k+1})^{2}$} (referred to \text{$J_{\alpha}$} in the following)  can be expressed as
\begin{equation}
\label{equ8}
J_{\alpha} = [i_{\alpha}^{*}-(1-\frac{R_{\rm g}T_{\rm s}}{L_{\rm g}})i_{\alpha}^{k}-\frac{T_{\rm s}}{L_{\rm g}}(v_{\alpha}^{k}-e_{\alpha}^{k})]^{2}.
\end{equation}

Through manipulation, the above item in (8) can be transformed into a new formulation as:
\begin{equation}
\label{equ9}
J_{\alpha}=\underbrace{(\frac{V_{\rm dc}T_{\rm s}}{2L_{\rm g}})^{2}}_{k_{\rm v}}[\underbrace{\frac{L_{\rm g}}{T_{\rm s}}i_{\alpha}^{*}-(\frac{L_{\rm g}}{T_{\rm s}}-R_{\rm g})i_{\alpha}^{k}+e_{\alpha}^{k}}_{\zeta_{\alpha}^{*}:\ \ \text{virtual error}}-u_{\alpha}^{k}]^{2}.
\end{equation}

The derivation in the second item \text{$J_{\beta}$} is the same. Owing to the item \text{$J_{\beta}$} having the same coefficient \text{$k_{\rm v}$}, the influence in the cost function is equal; thereby, it can be eliminated during cost function design, i.e., revised cost function (7) is derived.

\textit{Remark \uppercase\expandafter{\romannumeral1}:}
This transformation step is significant. On one hand, it enhances the universality of the mapping laws for different systems. As shown in (7), after the transformation, the selection of the optimal vector is only related to the virtual errors. This implies that even for different control objectives such as power, current, torque/flux, etc., after being transformed into virtual errors, the optimal vector can be obtained through the same mapping logic. On the other hand, this transformation also facilitates the fitting of the subsequent logical mapping relationship. Because, after the transformation, the fitting of the mapping laws is only related to the virtual errors, avoiding the influence of parameter changes such as discretization time, inductance, and resistance values on the mapping relationship\footnote{Note that this do not mean that the method is robust to the system parameter mismatch, which is a normal questions for model predictive control. Through the virtual error transformation the influence of parameter mismatch is reflected on the mismatched virtual error [see, (9)]. Thereby, additional process can be integrated to correct the calculation of virtual error, while no change will happen in the mapping laws.}.

\subsection{Mapping Logic Generation}
To select the optimal vector according to the established virtual errors, the mapping laws need to be fitted. In this time, two typical scenarios will be discussed: 1) basic case (with the single target to track reference currents); 2) general case (considering the penalty in switching frequency). 

\subsubsection{Basic case}

To make it easy for understanding, a basic fitting case is first discussed here. As derived in Sec.~\uppercase\expandafter{\romannumeral3}. A, the cost function is transformed into a formulation with virtual errors. Thereby, the fitting is separated from the entire system, which is simple and straightforward, generally containing three steps: 1) assume the possible range of the virtual errors. 2) deploy the conventional enumeration process to obtain the optimal vector. 3) logic boundaries derivation.

\begin{algorithm}[!h]
\caption{Logic Fitting Procedure (simple case)}
\label{A1}
\renewcommand{\algorithmicrequire}{\textbf{Input:}}
\renewcommand{\algorithmicensure}{\textbf{Output:}}
\begin{algorithmic}[1]
        \REQUIRE $\zeta_{\alpha\beta}^{*}$ in designed range (precise analysis or trial)   %%input
        \ENSURE Mapping laws (linear boundaries, i.e., $y=kx+b$)   %%output
        \FOR{$\zeta_{\alpha\beta}^{*}$ in range}
        \FOR{$\bm{u}_{\rm abc} \in\mathcal{U}_{8}$}
         \STATE $\bm{u}_{\alpha\beta} = \mathbf{M}_{\mathrm{C}}\bm{u}_{\rm abc}$
         \STATE Calculate $J_\mathrm{g}= \| {\zeta}_{\alpha}^{*} - {u}_{\alpha} \|^{2} + \| {\zeta}_{\beta}^{*} - {u}_{\beta} \|^{2}$
         \STATE Record $\bm{u}_{\rm abc}$ with minimal $J_\mathrm{g}$
        \ENDFOR
            \STATE Collect data $\Rightarrow$ each $\zeta_{\alpha\beta}^{*}$ with one optimal vector $\bm{u}_{\rm abc}^{\rm best}$
        \ENDFOR
        \STATE Calculate boundaries through least squares algorithm
        \RETURN Mapping laws (boundaries) $[k_{1},b_{1}],[k_{2},b_{2}],\dots$
    \end{algorithmic}
\end{algorithm}

The range assumption can be determined precisely according to the analysis for the possible operating ranges for the converter, or just ''cut-and-trial" method (which is simple and effective). Subsequently, the possible virtual errors are input to the cost function with the enumeration algorithm to obtain the optimal switching combination. The result is illustrated in Fig.~\ref{fig_5}. The horizontal and vertical coordinates denote virtual errors \text{$\zeta_{\alpha}^{*},\zeta_{\beta}^{*}$}, with color indicating the optimal switching vector for the current state, e.g., as the virtual error is located in the green region, the optimal switching vector is identified \text{$[0,1,0]$}. As shown, the boundaries of each switching combination are linear, which can be derived by the least squares algorithm. A pseudo-code fitting procedure for the mapping laws is provided as Algorithm 1.

\textit{Remark \uppercase\expandafter{\romannumeral2}:} Different from intelligent fitting methods such as artificial neural networks with uncertainty structures and overfitting problems, the logical maps in this method are certain and visualized, which can be directly described as a linear function, more reliable and simple to implement.

\begin{figure}[!tbp]
	\centering
	\includegraphics[width=3.4in]{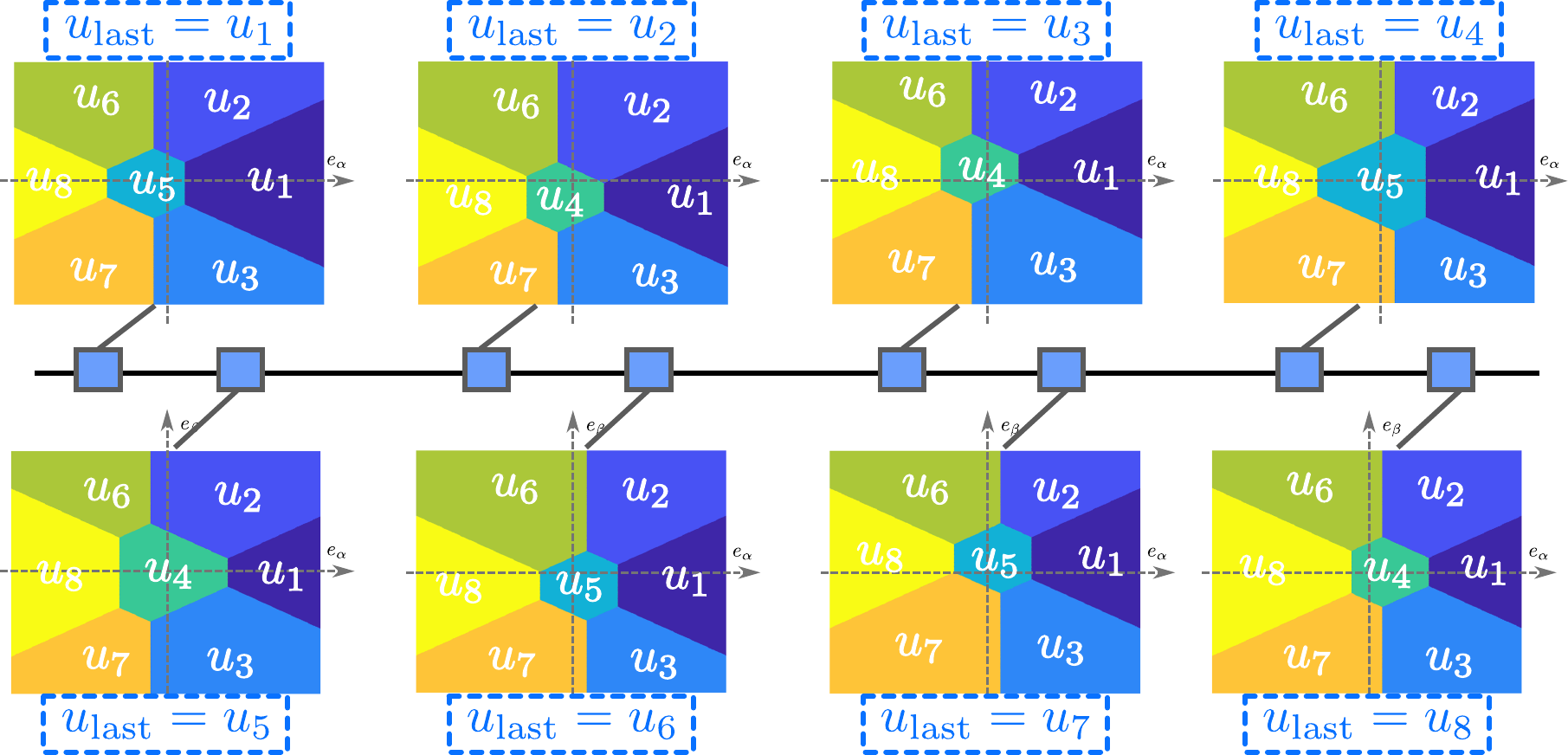}
	\caption{Mapping laws considering switching penalty (\text{$\lambda_{\rm g}^{\rm sf}=0.5$}).}
	\label{fig_6}
     \vspace{-5mm}
\end{figure}

\begin{algorithm}[!b]
\caption{Logic Fitting Procedure (general case)}
\label{A2}
\renewcommand{\algorithmicrequire}{\textbf{Input:}}
\renewcommand{\algorithmicensure}{\textbf{Output:}}
\begin{algorithmic}[1]
        \REQUIRE $\zeta_{\alpha\beta}^{*}$ in designed range, $\lambda_\mathrm{g}^\mathrm{sf}$ penalty factors   %%input
        \ENSURE Mapping laws (linear boundaries, i.e., $y=kx+b$)   %%output
        \FOR{$\zeta_{\alpha\beta}^{*}$ in range}
        \FOR{$\bm{u}_{\rm abc}^{k-1} \in\mathcal{U}_{8}$}
        \FOR{$\bm{u}_{\rm abc} \in\mathcal{U}_{8}$}
        \STATE $\bm{u}_{\alpha\beta} = \mathbf{M}_{\mathrm{C}}\bm{u}_{\rm abc}$
         \STATE $J_\mathrm{g}= \| \bm{\zeta}_{\alpha\beta}^{*} - \bm{u}_{\alpha\beta} \|^{2} + {\lambda}_\mathrm{g}^\mathrm{sf} \| \bm{u}_{abc}^{k-1} - \bm{u}_{abc} \|^{2}$
         \STATE Record $\bm{u}_{\rm abc}$ with minimal $J_\mathrm{g}$
        \ENDFOR
            \STATE Collect data $\Rightarrow$ each $(\zeta_{\alpha\beta}^{*}, \bm{u}_{\rm abc}^{k-1})$ with one $\bm{u}_{\rm abc}^{\rm best}$
        \ENDFOR    
        \ENDFOR
        \STATE Calculate boundaries through least squares algorithm
        \RETURN Mapping laws $[k_{1,1},b_{1,1}],[k_{1,2},b_{1,2}],\dots$
    \end{algorithmic}
\end{algorithm}

\subsubsection{General case}
Through the above simple case, the methods can realize the basic control objectives in several main power electronic products, e.g., power for grid-tied converters, flux/torque for motor drives. However, besides the basic control targets, the switching frequency is also considered as an optimal item in the control under the above scenarios. Therefore, the general cost function is designed as

\begin{equation}
\label{equ10}
\begin{aligned}
    J_\mathrm{g}(\bm{\zeta}_{\alpha\beta}^{*}, \bm{u}_{abc}^{k-1})= \underbrace{\| \bm{\zeta}_{\alpha\beta}^{*} - \bm{u}_{\alpha\beta}^{k} \|^{2}}_{\text{virtual item}} + {\lambda}_\mathrm{g}^\mathrm{sf} \underbrace{\| \bm{u}_{abc}^{k-1} - \bm{u}_{abc}^{k} \|^{2} }_{\text{switching item}},
\end{aligned}
\end{equation}
where \text{$\lambda_\mathrm{g}^\mathrm{sf}$} is the factor penalizing the switching frequency. 

The derivation of the general mapping laws is similar to the discussion in the above simple case. However, due to the impact of the switching item, the fitting process needs to consider the last switching state \text{$\bm{u}_{\rm abc}^{k-1}$}; different last switching states will cause different mapping logic. The fitting result is illustrated in Fig.~\ref{fig_6}. As can be seen, the centers of different switching states (\text{$\bm{u}_{\rm abc}^{k-1}$}) are drifted due to the impact of the switching penalty. A pseudo-code fitting procedure for the mapping laws of genera is provided as Algorithm 2.

\begin{figure}[!tbp]
	\centering
	\includegraphics[width=3in]{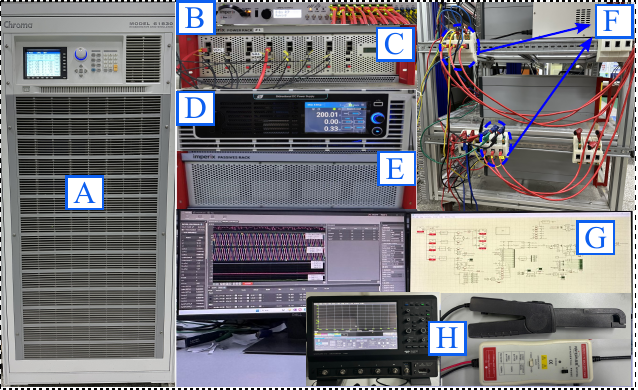}
	\caption{The lab-constructed experimental test bench. (a) Grid emulator. (b) Control platform B-BOX RCP. (c) 2L power converter PEN8024. (d) DC supply. (e) Three phase L filter. (f) current and voltage sensors. (g) PC. (h) Oscilloscope and probes.}
	\label{fig_7}
     \vspace{-5mm}
\end{figure}

\textit{Remark \uppercase\expandafter{\romannumeral3}:}
As depicted in Fig.~\ref{fig_6}, the mapping law varies with different switching situations, which means we need to store eight mapping laws for the DM-MPC under the 2-L topology. It is worth mentioning that each mapping law corresponds to a float-type array of twenty elements in the procedure; therefore, the total storage requirement for the eight laws is only around 0.625 KB, which can be considered negligible. For instance, the smallest flash memory offered by mainstream microcontroller series produced by ST is 16 kB, meaning that 0.625 KB accounts for only about 4\%.

\begin{table}[!hp]
	\caption{SYSTEM PARAMETERS DURING THE TEST}
	\centering
        \renewcommand{\arraystretch}{1}
        \scalebox{1.2}{
	\begin{tabular}{l c c r}
		%\toprule[1.5pt]
		\hline\hline 
		 Description & Parameter & Value & Unites  \\
		%\midrule[1pt]
		\hline
             Grid frequency &\text{$f_{\mathrm{g}}$}   & 50     &[\text{Hz}]  \\
             DC Voltage &\text{$V_\mathrm{dc}$}     & 150     &[\text{V}]  \\
             Grid Voltage &\text{$V_\mathrm{ac}$}     & 50     &[\text{V}] \\
		 Filter resistance& \text{$R_\mathrm{g}$} & 0.044    &[\text{$\Omega$}]  \\
             Filter inductance& \text{$L_\mathrm{g}$}  & 5    &[\text{mH}]     \\
             Control period& \text{$T_\mathrm{s}$}     & 50    &[\text{us}]     \\
		\hline
		%\bottomrule[1.5pt]
	\end{tabular}
        }
\end{table}

\subsection{Implementation on Micro-processors}
Since the main computation is transferred to offline, the online deployment is quite efficient compared to FCS-MPC. The overall implementation is divided into four cascaded steps: 1) one-step prediction \text{$\bm{i}^{k+1}$} to compensate for the computational delay, 2) virtual error \text{$\bm{\zeta}^{*}$} calculation according to the measurement data (\text{$V_{\rm dc}^{k}$}, \text{$\bm{e}^{k}$}, etc.), 3) select the mapping law through the current switching state \text{$\bf{u}_{\rm abc}^{k}$}, 4) directly determine the optimal switching vector according to pre-stored laws.

\section{Experimental Verification}
To validate the effectiveness, a grid-tied two-level power converter prototype was constructed in the laboratory, which is shown in Fig.~\ref{fig_7}. The parameters are collected in Table \uppercase\expandafter{\romannumeral1}. 
%Detailed analyses are as follows.

\begin{figure}[htbp]
\centering
\subfloat[Inverter Mode: (\text{$i_{\rm d}=5$A} \text{$\Leftrightarrow$} \text{$i_{\rm d}=10$A} \text{$\Leftrightarrow$} \text{$i_{\rm d}=15$A})]
{
    \begin{minipage}[b]{1\linewidth}
        \centering
        \hspace{-5mm}
        \includegraphics[scale=0.69]{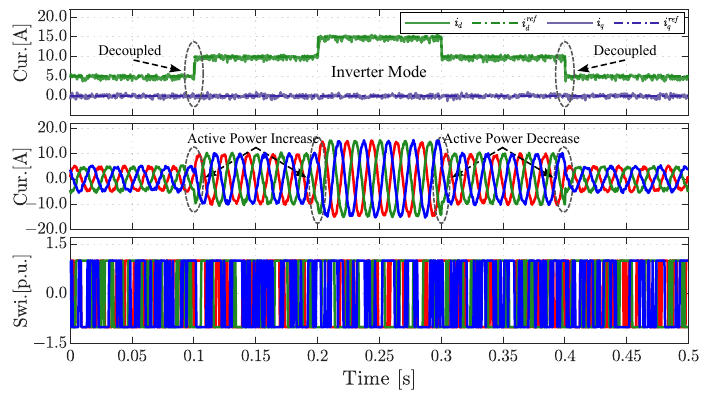}
    \end{minipage}
}\\
\subfloat[Rectifier Mode: (\text{$i_{\rm d}=-5$A} \text{$\Leftrightarrow$} \text{$i_{\rm d}=-10$A} \text{$\Leftrightarrow$} \text{$i_{\rm d}=-15$A})]
{
 	\begin{minipage}[b]{1\linewidth}
        \centering
        \hspace{-5mm}
        \includegraphics[scale=0.69]{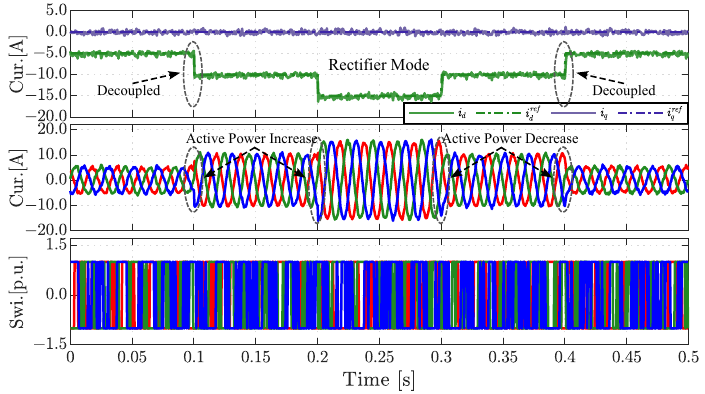}
    \end{minipage}
}\\
\subfloat[Mode Switching: (Inverter Mode \text{$\Leftrightarrow$} Rectifier Mode)]
{
 	\begin{minipage}[b]{1\linewidth}
        \centering
        \hspace{-5mm}
        \includegraphics[scale=0.69]{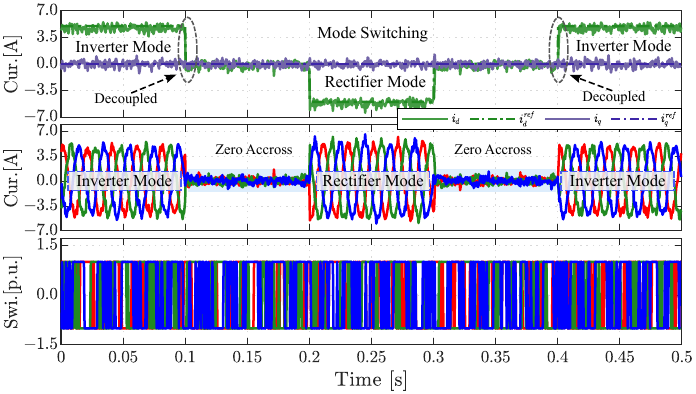}
    \end{minipage}
}
\caption{[Experimental Results] Overall assessment of the proposal.}
 \label{fig_8}   
  \vspace{-5mm}
\end{figure}

\subsection{Overall Assessment}

To validate the effectiveness, an overall assessment is firstly conducted, which includes inverter mode, rectifier mode, and the test under mode switching. The results are depicted in Fig.~\ref{fig_8}. As shown, the DM-MPC achieves flexible power control capability no matter under which scenario. In detail, fast current/power response can be achieved under transient while a precise trajectory tracing is obtained in steady-state. Additionally, the active power and reactive power are greatly decoupled during the current step.

\subsection{Consistency Verification}

\begin{figure*}[!tbp]
	\centering
        \hspace{-7mm}
	\includegraphics[width=7in]{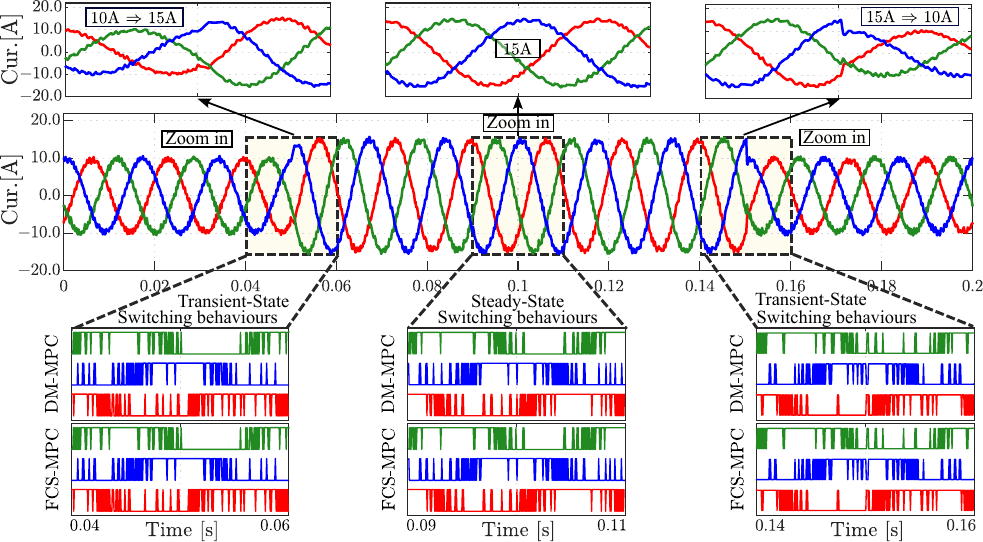}
	\caption{[Experimental Results] Consistency validation of the switching vector selection between the proposed DM-MPC and classical FCS-MPC.}
	\label{fig_9}
    \vspace{-5mm}
\end{figure*}

To validate the consistent control characteristic of the proposed DM-MPC and FCS-MPC, in this section, contrast tests are carried out at our laboratory-constructed test bench. During the test, both methods, i.e., the proposed and classical FCS-MPC, are downloaded to a single controller (B-BOX RCP, produced by Imperix) at the same time. Therefore, strictly consistent data can be input to both methods. The tests focus on two typical situations: 1) \textit{steady-state}; 2) \textit{transient-state}. The results are shown in Fig.~\ref{fig_9}. As depicted, during the experimental test, the DM-MPC selected completely the same switching combination as the classical FCS-MPC, regardless of the conditions, indicating that the computation reduction is no sacrifice in control.

\subsection{Switching Frequency Penalty}

\begin{figure}[!tbp]
\centering
\subfloat[DM-MPC: 15A, \text{$\lambda = 0.5$}]
{
    \begin{minipage}[b]{0.45\linewidth}
        \centering
        \hspace{-5mm}
        \includegraphics[scale=0.8]{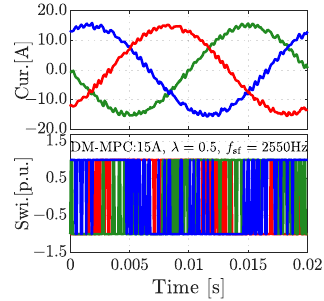}
    \end{minipage}
}
\subfloat[DM-MPC: 15A, \text{$\lambda = 0$}]
{
    \begin{minipage}[b]{0.45\linewidth}
        \centering
        \hspace{-5mm}
        \includegraphics[scale=0.8]{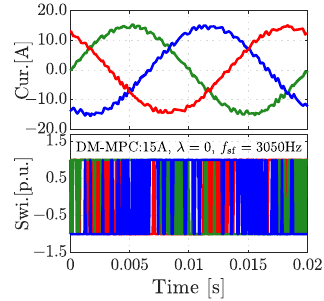}
    \end{minipage}
}\\
\subfloat[DM-MPC: 10A, \text{$\lambda = 0.5$}]
{
    \begin{minipage}[b]{0.45\linewidth}
        \centering
        \hspace{-5mm}
        \includegraphics[scale=0.8]{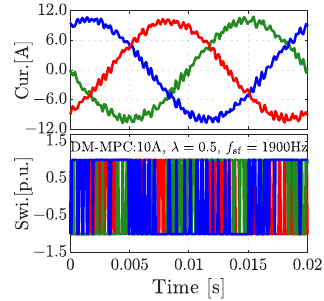}
    \end{minipage}
}
\subfloat[DM-MPC: 10A, \text{$\lambda = 0$}]
{
    \begin{minipage}[b]{0.45\linewidth}
        \centering
        \hspace{-5mm}
        \includegraphics[scale=0.8]{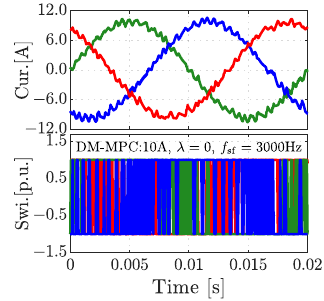}
    \end{minipage}
}
\caption{[Experimental Results] Switching frequency suppression.}
\label{fig_10}
\vspace{-5mm}
\end{figure}

Switching frequency is a crucial index for a control scheme, as it ensures that the system's heat loss does not exceed acceptable limits. In this section, the DM-MPC is given to different switching penalty factors \text{$\lambda_{\rm g}^{\rm sf}$} to validate its effectiveness in switching penalty. As illustrated in Fig.~\ref{fig_10}, the switching frequency with \text{$\lambda_{\rm g}^{\rm sf} = 0.5$} is 30\% lower compared to zero penalty. However, the penalty for switching frequency is not constant. With the variation of current amplitude, the switching frequency is variable. This is also a shortcoming of classical FCS-MPC. However, several methods are proposed to solve this problem, and the integration in this framework may be further explored.

\begin{figure}[!tp]
	\centering
	\includegraphics[width=3.3in]{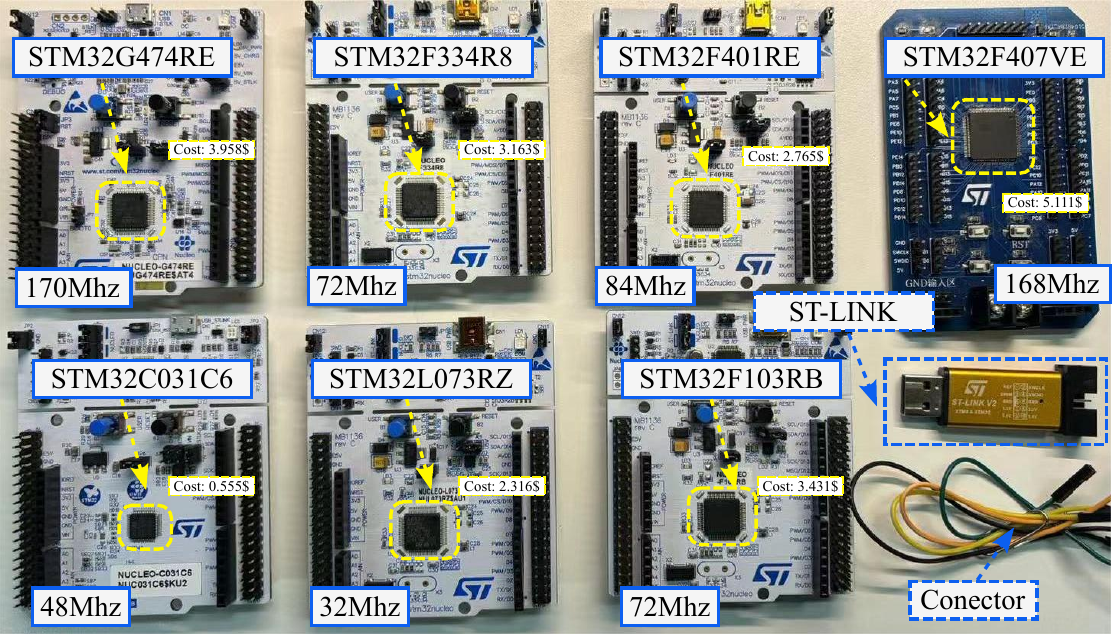}
	\caption{The setup of microprocessors in the computational analysis.}
	\label{fig_11}
    \vspace{-5mm}
\end{figure}

\subsection{Computation Analysis}

\begin{figure}[!htbp]
	\centering
        \hspace{-7mm}
	\includegraphics[width=3.1in]{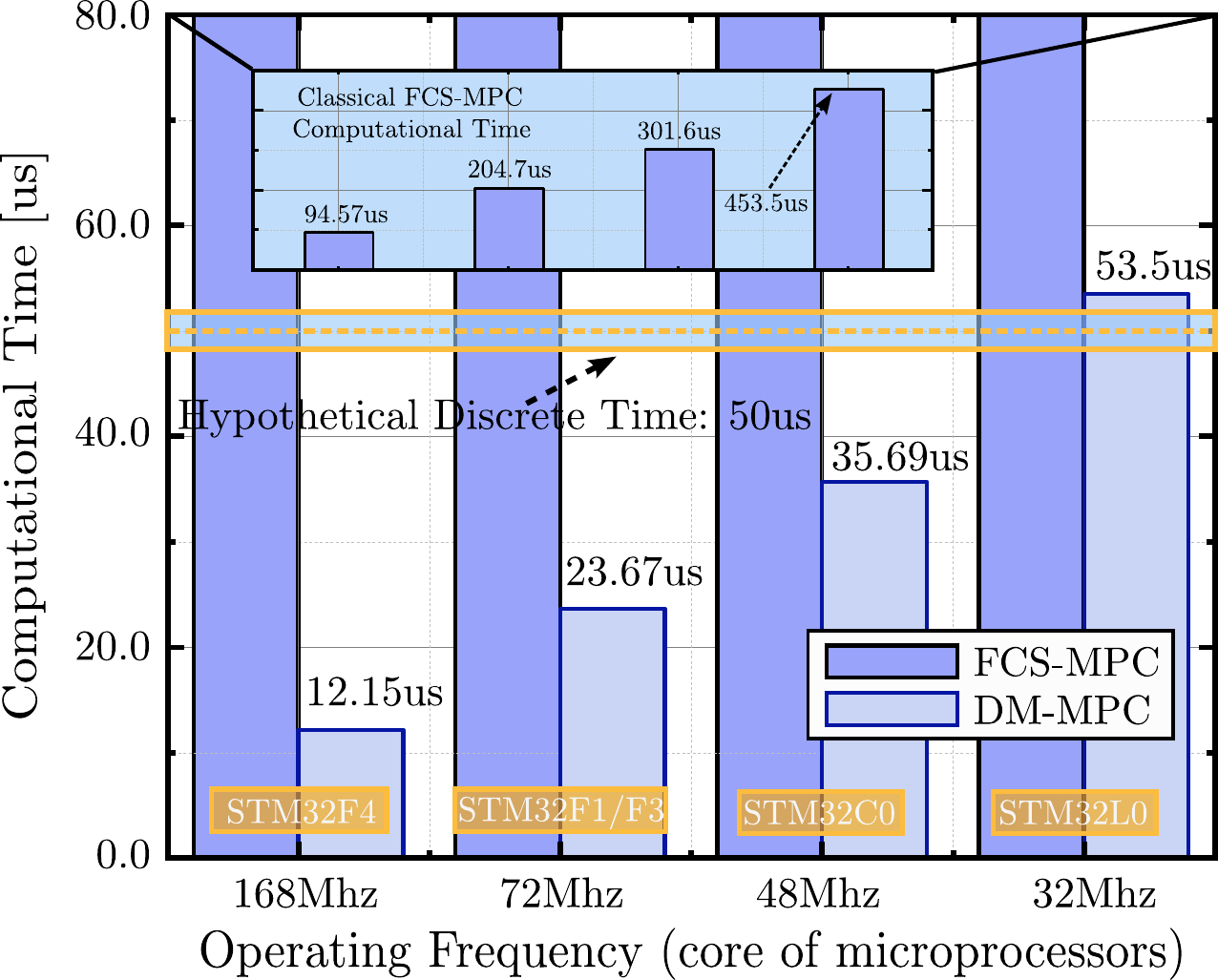}
	\caption{Computational analysis by tracing tools in MDK-ARM.}
	\label{fig_12}
    \vspace{-3mm}
\end{figure}

In this subsection, the computational load of the proposal is analyzed with a set of competitive microprocessors produced by ST, i.e., STM32G474RE, STM32F407VE, STM32F401RE, STM32F334R8, STM32F103R8 (see, Fig.~\ref{fig_11}). To ensure precision in the timing of the deployment procedure, the trace tool in MDK-ARM is utilized to track the computational time during the tests. Furthermore, to avoid deviations in computational time caused by different input data \text{$\bm{\zeta}^{*}, \bm{u}_{abc}^{k-1}, \bm{u}_{abc}^{k}$}, the same operating instant was used across the compared processors. The corresponding results are illustrated in Fig.\ref{fig_12}.

The operation frequency has a significant impact on the computational speed of micro-processors. Since the STM32C031C6 and STM32L073RZ series do not support the tracing function in MDK-ARM, the STM32G474 (which has the highest maximum operating frequency of 170 MHz among the tested processors) is utilized to simulate instead, which can be realized through the official tools STM32CubeMX. The results are shown in Fig.~\ref{fig_12}. 
In the figure, 168 MHz is the maximum operating frequency of the STM32F407, while 72 MHz is the maximum value for both the STM32F103 and STM32F334. The same logic applies to the others. The results demonstrate that the proposal significantly reduces the computational burden, e.g., the MPC with a desired control period of 50 µs deployed on the STM32C031 is reduced from 301.6 µs to 35.69 µs.

To quantify the degree of optimization, a reduction factor is designed:
\begin{equation}
\label{equ12}
\eta_{\rm co} = \frac{T_{\rm fcs}-T_{\rm dm}}{T_{\rm fcs}} \times 100\% .
\end{equation}

The experimental data are generally collected in Table~\ref{table1}. As calculated, the DM-MPC can reduce the computational burden by 87\% to 88\% compared to the classical FCS-MPC.

\section{Conclusion}
In this article, we propose a direct mapping model predictive control strategy (DM-MPC) to reduce the considerable computational load in classical finite control set based model predictive control (FCS-MPC). Through designing the visual mapping laws, the DM-MPC can directly select the optimal voltage vector without any enumeration process. Experimental results demonstrate that DM-MPC achieves the completely same control performance as FCS-MPC while requiring only 12\% of the computational resources compared to the classical approach. The proposed DM-MPC method is characterized by the following features and advantages.
\begin{enumerate}
    \item It is simple and effective, resulting in an easy solution for industrial products.
    \item It can obtain the same control performance during implementation as the classical FCS-MPC.
    \item It offers extremely computational efficiency (the reduction of computational burden is up to \textit{88\%})
\end{enumerate}

Future work will focus on extending this approach to multiple applications.

\begin{table}[!t]
\setlength{\tabcolsep}{6pt}
\renewcommand\arraystretch{1.6}
	\caption{Computational Analysis in \\ Several Typical Micro-possessors Produced by ST\label{table1}}
	\centering
	\begin{tabular}{|c|c|c|c|c|}
     \hline
     \rowcolor{mycolor_c2}{\diagbox{Type}{Data}} & \makecell[c]{Time\\DM-MPC} &\makecell[c]{Time\\FCS-MPC} & \makecell[c]{Operation\\Frequency} & \makecell[c]{Reduced\\
     burden \text{$\eta_{\rm co}$}} \\
     \hline
     \makecell[c]{STM32G474RE}
     &\makecell[c]{12.2 us} &  {94.5 us}
     & {170 MHz} & {87.1\%}\\
     \hline
     \makecell[c]{STM32F407VE}
     & \makecell[c]{14.3 us} &  \makecell[c]{120.5 us}
     & {168 MHz} & {88.1\%}\\
     \hline
     \makecell[c]{STM32F401RE}
     & {20.4 us} &  \makecell[c]{175.6 us}
     & \makecell[c]{84 MHz} & {88.4\%}\\
     \hline
     \makecell[c]{STM32F334R8}
     & \makecell[c]{33.2 us} &  {285.2 us}
     & \makecell[c]{72Mhz} & {88.4\%}\\
     \hline
    \makecell[c]{STMF103RB}
     & \makecell[c]{31.6 us} &  \makecell[c]{275.8 us}
     & {72MHz} & {88.5\%}\\
     \hline
	
	\end{tabular}
\end{table}

\balance
\bibliographystyle{IEEEtran}
\bibliography{lib}

\end{document}